\documentclass[letterpaper, 10 pt, conference]{ieeeconf}  

\IEEEoverridecommandlockouts                              

\usepackage{cite}
\usepackage{amsmath,amssymb,amsfonts}
\usepackage{algorithmic}
\usepackage{graphicx}
\usepackage{textcomp}
\usepackage{xcolor}
\usepackage{float}
\usepackage{booktabs}
\usepackage{gensymb}
\usepackage{float}
\usepackage{caption} 
\usepackage{afterpage} 
\usepackage[acronym]{glossaries}
\newacronym{ramis}{RAMIS}{robot-assisted minimally invasive surgery}
\newacronym{fov}{FOV}{field-of-view}
\newacronym{roi}{ROI}{region of interest}
\newacronym{dof}{DoF}{degrees of freedom}
\newacronym{gui}{GUI}{graphical user interface}
\usepackage[square, numbers, sort&compress]{natbib}
\usepackage{color,soul}

\title{\LARGE \bf Where is the Boundary: Multimodal Sensor Fusion Test Bench for Tissue Boundary Delineation}

\author{Zacharias Chen$^{*{\dagger}1,4}$, Alexa Cristelle Cahilig$^{*1,4}$, Sarah Dias$^{2}$, Prithu Kolar$^{3}$, Ravi Prakash$^{{\dagger}1,4}$, Patrick J. Codd$^{1}$ 
\thanks{$^{1}$ Duke University, NC, USA}
\thanks{$^{2}$ Research Triangle High School, NC, USA}
\thanks{$^{3}$ North Carolina School of Science and Mathematics, NC, USA}
\thanks{$^{4}$ Supported by the Duke Bass Connections Student Research Award}
\thanks{$*$ Both authors contributed equally}
\thanks{$^{\dagger}$Corresponding author: Zacharias Chen (zachariastchen@gmail.com) \& Ravi Prakash (ravi.prakash@duke.edu)
 }
}
\begin{document}
\maketitle
\begin{abstract}
    Robot-assisted neurological surgery is receiving growing interest due to the improved dexterity, precision, and control of surgical tools, which results in better patient outcomes. However, such systems often limit surgeons' natural sensory feedback, which is crucial in identifying tissues --- particularly in oncological procedures where distinguishing between healthy and tumorous tissue is vital. While imaging and force sensing have addressed the lack of sensory feedback, limited research has explored multimodal sensing options for accurate tissue boundary delineation. We present a user-friendly, modular test bench designed to evaluate and integrate complementary multimodal sensors for tissue identification. Our proposed system first uses vision-based guidance to estimate boundary locations with visual cues, which are then refined using data acquired by contact microphones and a force sensor. Real-time data acquisition and visualization are supported via an interactive graphical interface. Experimental results demonstrate that multimodal fusion significantly improves material classification accuracy. The platform provides a scalable hardware-software solution for exploring sensor fusion in surgical applications and demonstrates the potential of multimodal approaches in real-time tissue boundary delineation.
        
    
\end{abstract}

\section{Introduction}
 Approximately 13.8 million patients with neurological disorders require brain surgery \cite{Dewan2019-ec}. Recently, interest in robot-assisted surgery (RAS) for neurological applications has surged due to the demand for minimally invasive surgical techniques that offer increased precision and dexterity \cite{LIN2023e22523}, with improved post-operative recovery. However, because the robot serves as a proxy between the surgeon and the patient, RAS limits direct access to critical sensory feedback, such as tactile, visual, and auditory cues—that are essential for tasks like tissue boundary delineation. This sensory gap limits the adoption of RAS in neurosurgery. These challenges are not new, and different sensing and control schemes have been developed to compensate for the lack of natural sensing methods. Previous works have explored the use of vibro-acoustic \cite{10404852}, fluorescence \cite{9629801}, force \cite{10485247}, and image-based \cite{8513556} methods. Works such as Liao et al. \cite{LIAO2012754} developed a robotic laser ablation system guided by 3D Magnetic resonance imaging (MRI) and 5-aminolevulinic acid (ALA)-induced fluorescence for brain tumor diagnosis and removal. However, the use of multimodal sensing for real-time feedback is limited, especially for the task of delineating tissue boundaries, and presents a need for hardware-software co-design.

In this study, we present a multimodal test bench for quantifying the impact of sensing modalities on tissue boundary delineation. Our contributions are:
\begin{itemize}
    \setlength{\itemsep}{-1.5pt}
    \item A tissue boundary detection method leveraging the multi-modal information streams from visual, acoustic, and tactile sensing modalities. Here, a visual sensor helps guide the initial starting estimates, followed by vibration and force sensor data to further refine optically ambiguous regions.
    \item A user-friendly experiment setup allowing real-time control and visualization for investigating multimodal tissue boundary detection, given the noisy nature of tool-tissue interaction. The modular nature allows the implementation of novel algorithms and sensors with ease.
\end{itemize}

\begin{figure*}[!h]  
    \centering
    \includegraphics[width=0.95\textwidth]{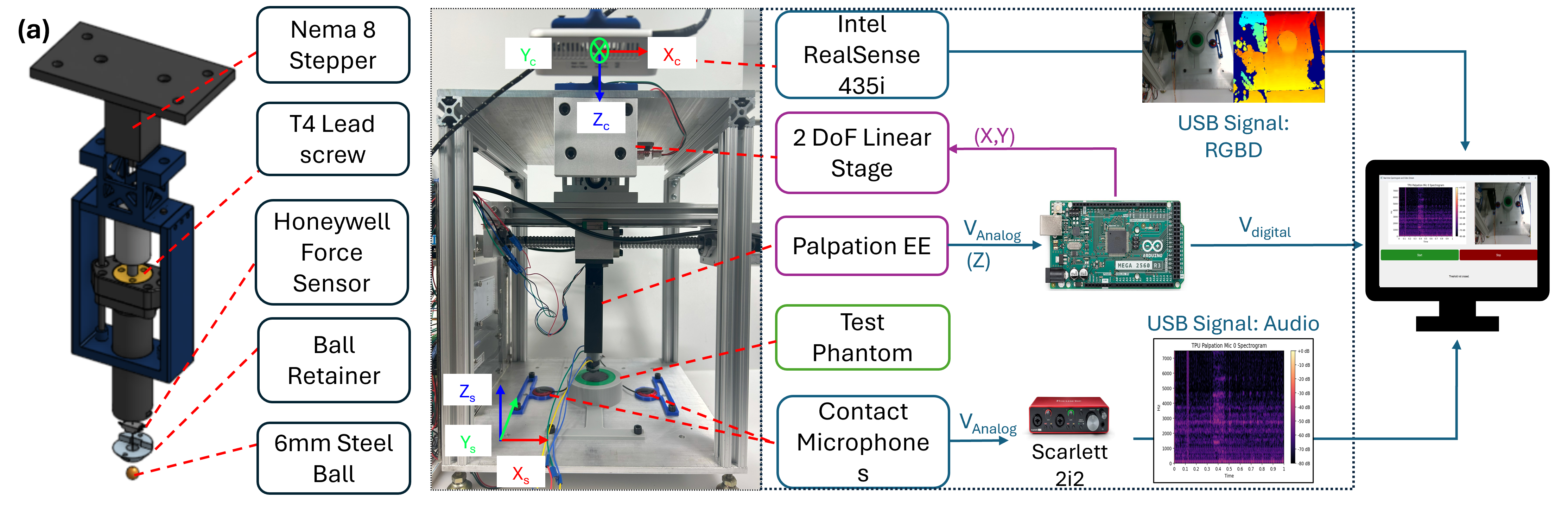}
    \caption{\textbf{System Overview.} (a) palpation end effector (b) test bench setup (c) system flow diagram (d) user interface }
    \label{fig:overview}
\end{figure*}


\section{Methods}

\subsection{System Overview}
The experimental system consists of a portable 2 \acrlong{dof} test bench with complementary sensing modalities providing local and global information coupled with an easy-to-use user interface for real-time data streaming and control. The \textit{(X,Y)} motion is achieved through two downwards mounted RATTMMOTOR ZBX80 200 mm linear stages [RATTMMOTOR, China]. A custom lead screw driven end effector, equipped with a Honeywell FSS005WNSB [Honeywell, NC, USA] force sensor, serves as the \textit{Z} axis and the force-sensing modality. A statically mounted Intel RealSense 435i [Intel, CA, USA] camera and two piezo contact microphones [TraderPlus] provide visual and auditory feedback, respectively. The fundamental functionalities associated with the force sensor and all actuators are controlled by an Arduino Mega 2560 V3 microcontroller, which communicates with a computer via a simple serial protocol over USB. A central python script handles the interleaving of stage commands with all sensor channels to allow for multimodal data collection and more complex functionalities, such as vision-guided boundary refinement. All of these functionalities are available to the user through a \acrlong{gui}.

\subsection{Data Acquisition}
The RGB and depth channels from the camera are streamed at a resolution of 640 x 480 pixels and 15 frames per second over USB. Each contact microphone is connected to an audio interface, Scarlett 2i2 [Focusrite, United Kingdom], to minimize latency and improve signal-to-noise ratio. The audio interface is connected to the computer via USB as well. Raw force sensor data are passed through an instrumentation amplifier with a 42.2 Hz low-pass filter and a 10-bit ADC to an Arduino Mega, where the voltage is calibrated to output values in Newtons (N). A multi-thread processing approach allows for minimal latency acquisition of the aforementioned data in real time, as seen in Figure \ref{fig:overview}.


\subsection{System Calibration}
Hand-eye calibration is performed in eye-to-hand configuration between the camera and the stage with an average Euclidean distance error of 1.04 ± 0.45 mm. A green laser probe was placed concentrically with the end-effector shaft to project the $(X_{stage}, Y_{stage})$ coordinates into the camera frame at $Z_{stage}$ levels: 0 mm, 15 mm, and 30 mm. The centroid of each laser point was segmented in $(X_{pixel}, Y_{pixel})$, and the corresponding depth at that point was retrieved. The $(X_{pixel}, Y_{pixel}, Z_{camera})$ point is then deprojected using the camera's intrinsics to yield $(X_{camera}, Y_{camera}, Z_{camera})$. Singular value decomposition was applied to calculate the best transformation between the two sets of points, shown in Equation \ref{eq:registration}. p, s, R, t represent the points, scaling, rotation, and translation, respectively. Depth readings fluctuated by ± 1mm over time, which can be attributed to the camera's inherent accuracy, as well as the distance and angle at which the camera was mounted. 



\begin{equation}
\mathbf{p}_{\text{stage}} =
\underbrace{
\begin{bmatrix}
\mathbf{s} \mathbf{R} & \mathbf{t} \\
\mathbf{0}^\top & 1
\end{bmatrix}
}_{\mathbf{T}_{\text{camera} \to \text{stage}}}
\mathbf{p}_{\text{camera}}
\label{eq:registration}
\end{equation}

\section{Experiments \& Results}
To demonstrate our system's ability to leverage multimodal sensing data for delineating tissue boundaries, we first verify the functionality of each modality in classifying materials through an image-guided data collection method on high-stiffness materials, as shown in Figure \ref{fig:overview} (a). This is followed by validating the method on challenging and realistic material, excised \textit{ex vivo} tissues. 

\subsection{Material Classification via Multi-sensor Feedback}
\begin{figure}[H]  
    \centering \includegraphics[width=0.95\columnwidth]{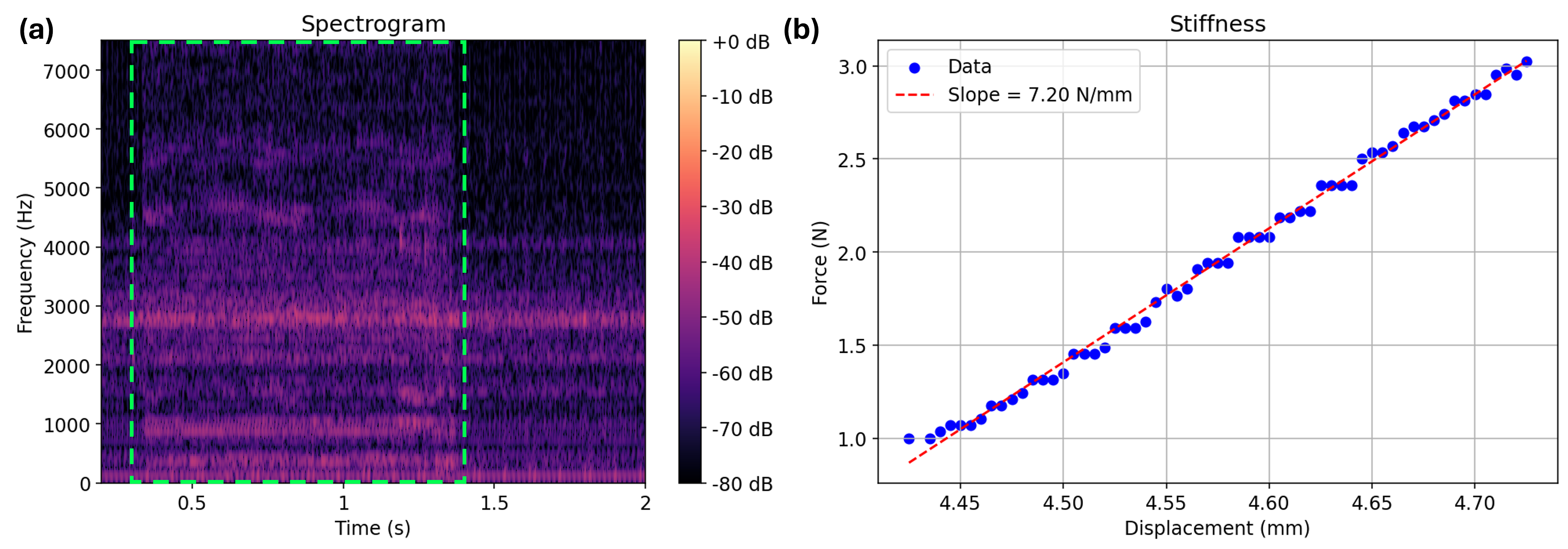}
    \caption{\textbf{Single Point Palpation Data on 5\% TPU}. (a) spectrogram and, (b) force vs. displacement plot for the palpation }
    \label{fig:dataGraphs}
\end{figure}

A system validation is performed on different 3D-printed cylindrical objects composed of polylactic acid (PLA) at 5\% and 15\% infill density, thermoplastic polyurethane (TPU) at 5\% infill density, and \textit{ex vivo} porcine muscle tissue. Simultaneous force and vibration data are collected by palpating the objects in a $10 \times 10$ raster pattern with a step size of 1 mm. Figure \ref{fig:dataGraphs} displays the spectrogram generated by one of the contact microphones as well as the force vs displacement data captured by the end-effector-mounted force sensor by palpating a single tissue point. The audio recording started shortly before the end-effector makes contact with the sample and terminated shortly after the palpation finished. Average stiffness of materials using force sensor is shown in Table \ref{tb:material stiffness}.

Material classification was performed using a support vector machine (SVM) and a simple multilayer perceptron (MLP) based on extracted features. The stiffness, indenter displacement, and smoothness of the raw data are used as characteristics of the force sensor, while the Mel-frequency cepstral coefficients (MFCC) of the order 12 are extracted from each contact microphone recording as a representative characteristic. 

\begin{figure*}[h!]  
    \centering
    \includegraphics[width=0.95\textwidth]{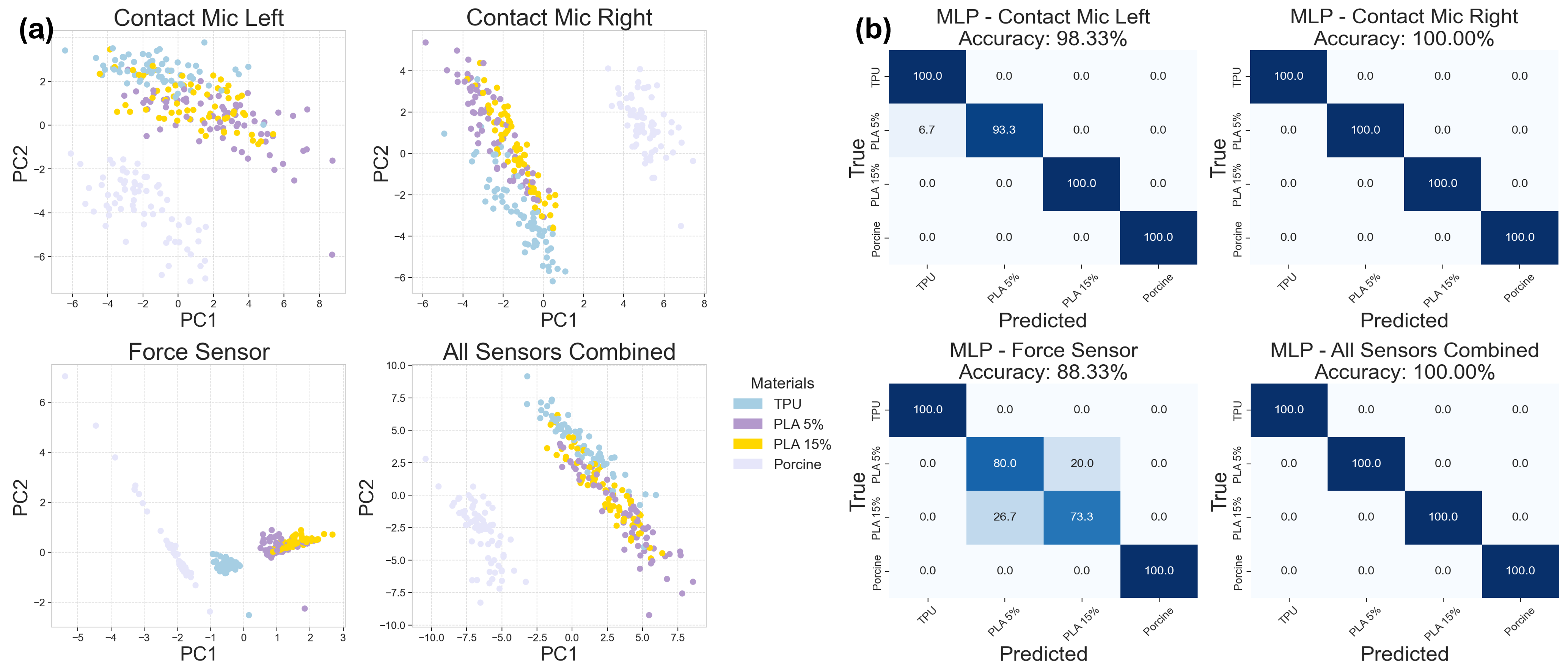}
    \caption{\textbf{Classification Results.} (a) PCA analysis (b) MLP confusion matrices for contact microphones, force sensor, and all sensors combined} 
    \label{fig:Classification}
\end{figure*}
Principal component analysis (PCA) of the raw features highlights the ability of the force sensor to capture differences in object properties better than contact microphones, as seen in Figure \ref{fig:Classification} (a). PLA, TPU, and porcine form distinct clusters under the two principal components of the force sensor, but the plastics exhibit grouping in the two vibrational contact microphone sensors. The results show a generally improved performance by MLP over SVM with all of the sensors combined, with MLP achieving 100\% accuracy on the testing dataset while SVM achieved 92\%. Within sensors, the classifiers show misclassification between PLAs, especially when only utilizing the force sensor, but combining features from all the sensors shows improved results as seen in Figure \ref{fig:Classification} (b). Misclassifications are observed between TPU and 5\% PLA in the left contact microphone MLP classifier.  The MLP classifiers displayed 88.33\% accuracy when solely relying on the force sensor, showing misclassifications between 5\% infill and 15\% infill PLA.

\vspace{-.5em}    
\begin{table}[H]
\centering
\caption{Average Stiffness of Materials (N/mm ± SD)}
\label{tb:material stiffness}
\begin{tabular}{|l|l|}
\hline
\textbf{Material} & \shortstack{\textbf{Average Stiffness} $\pm$ \textbf{Std Dev } }\\
\hline
PLA 15\%& 30.3875 $\pm$ 0.2283 \\
\hline
PLA 5\% & 23.7667 $\pm$ 0.2767 \\ \hline
TPU & 7.8982 $\pm$ 0.2870 \\
\hline
Porcine & 0.3286 $\pm$ 0.0343 \\
\hline
\end{tabular}
\vspace*{-0.7\baselineskip}
\end{table}

\subsection{Vision Guided Boundary Search}
\begin{figure}[h!]  
    \centering \includegraphics[width=0.95\columnwidth]{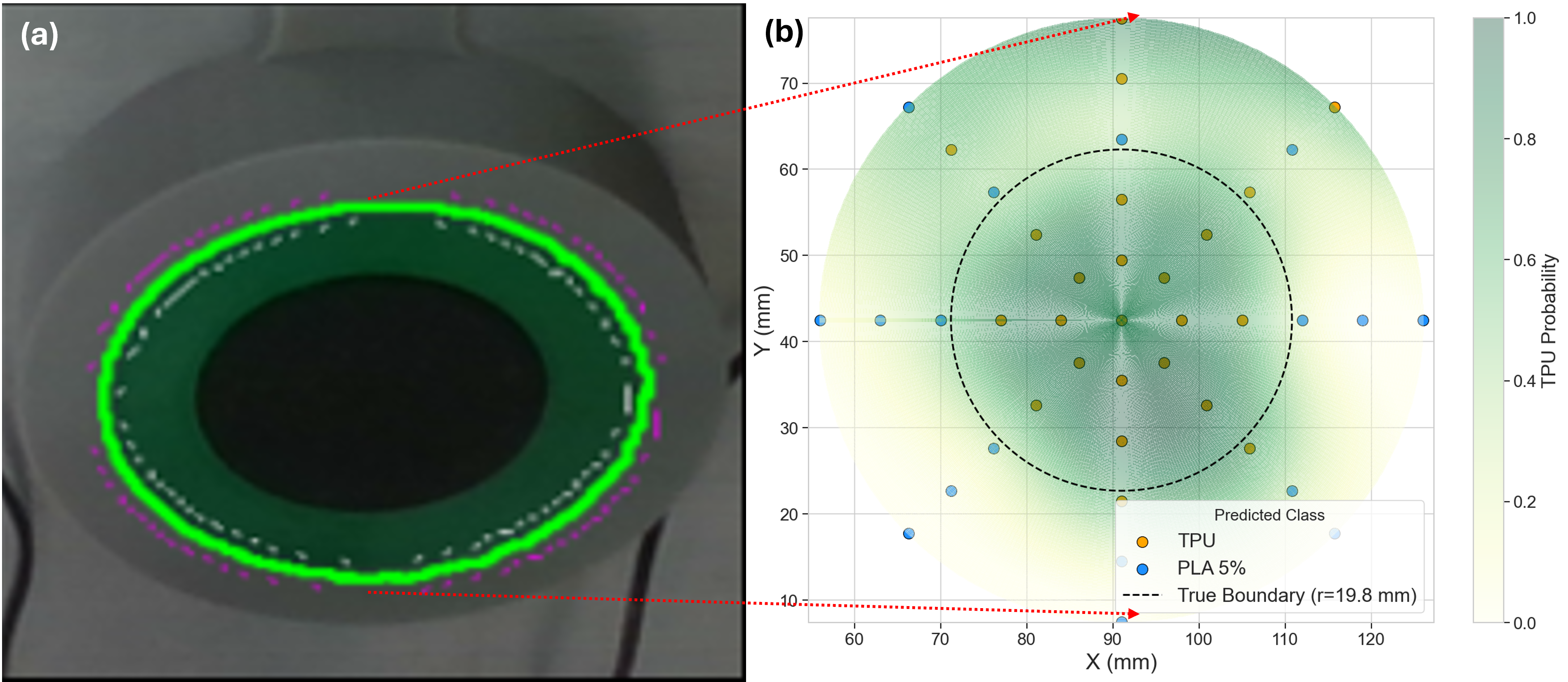}
    \caption{\textbf{Vision Guided Boundary Search}. (a) user designated region of interest (b) TPU classification and probability map}
    \label{fig:Vision Guided}
    \vspace{-0.5em}    
\end{figure}

To evaluate the performance of the classifier for boundary detection, three 3D-printed cylinders consisting of black TPU at 5\% infill density, green PLA at 5\% infill density, and gray PLA at 15\% infill density are placed in a concentric stack. Using the camera feed, the user selects an initial region of interest to be investigated. This generates a set of coordinate points that originate from the centroid of the region and move radially outward in spokes. Force and vibration data are collected simultaneously at each of the probed points. Figure \ref{fig:Vision Guided} displays the selected region within the GUI and the resulting probability map from the SVM classifier with radial interpolation. The prediction results are consistent with ground truth within the tissue boundary though, outliers can be observed outside the boundary.

\subsection{Tissue Classification}
\begin{figure}[h!]  
    \centering \includegraphics[width=0.95\columnwidth]{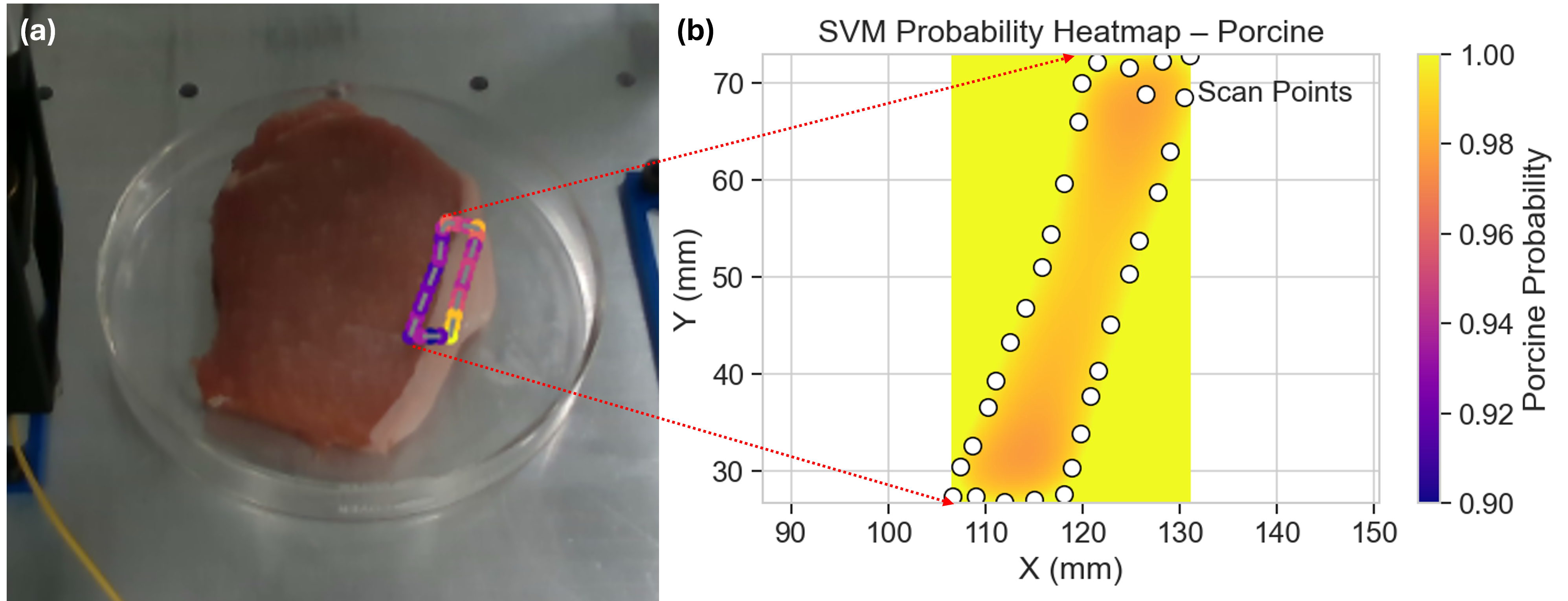}
    \caption{\textbf{Tissue Boundary Characterization}. (a) user-defined boundary (b) porcine muscle probability map (SVM)}
    \label{fig:Tissue}
    \vspace{-0.5em}    
\end{figure}

The methods developed for the interrogation of hard materials are extended to soft tissues. An \textit{ex vivo} piece of porcine tissue is placed in a Petri dish and multimodal data are collected following a user-defined polyline that overlaid the boundary between muscle tissue and fat. The resulting classification probability of the tissue being muscle is determined through the SVM trained on all sensors and shown in Figure \ref{fig:Tissue}. SVM was chosen for its superior experimental performance in soft tissues. 

\section{Discussion \& Conclusion}
The proposed test bench reliably delineates tissue boundaries via multimodal fusion and both guided and blind exploration.
The PCA reveals clear clusters between the stiffness response of the materials, although PLA features between the two infill densities are seen to be grouped close to each other in PCA analysis. This aligns with known material properties, since PLA, TPU, and porcine have relatively distinct stiffness levels that outweigh the variability in stiffness afforded by a marginal change in infill density. 

Improved classification accuracy achieved through the integration of both tactile and vibration sensors underscores the importance of multimodal sensing. Force sensors primarily capture local mechanical properties such as stiffness and elastic modulus. In contrast, vibration sensors (contact microphones) detect resonance information not only from the tissue under the end-effector but also from neighboring materials, including elasticity-related features. The ability to capture a wider spread of tissue material gives vibrational sensors the potential to capture features that may be essential for tissue delineation but are missed by force sensors alone. However, the widespread nature of vibrational data can also be a drawback, particularly near the junction of dissimilar tissues, where the signals may no longer significantly represent the local tissue of interest but majorly information from adjacent materials. 

This is later reflected in the performance of the multimodal classifiers in identifying the true boundaries, as shown in Figure \ref{fig:Vision Guided} b. In particular, there are few discrepancies between the predicted and ground truth classification of data points near the boundary, as TPU extends radially towards PLA 5\%, as seen in Figure \ref{fig:Vision Guided}. These discrepancies are further exacerbated by the slow contact speed of the lead-screw end-effector, generating a relatively weak audio signature upon initial contact. Under such conditions, high-variation components can be masked under systematic noise and not be accounted for when training the MLP.


\section{Future Work}
To address the aforementioned observations, future work will focus on engineering an end effector that is capable of generating stronger audio signatures while improving the force range and resolution to better suit soft tissue characterization \cite{prakash2023brain}, which often exhibits elastic behavior in the milli-Newton range. In general, a better representative selection of audio features is needed, and given the improved results of the MLP classifier, a spectrogram-based feature extractor seems promising and would be used together with a convolutional neural network. Additionally, active search algorithms \cite{9698135} that weigh high-probability boundary frontiers and advanced classification algorithms will be the areas of exploration in the future.

\section{Acknowledgments}
The authors wish to thank Kent Yamamoto and the members of the Brain Tool Lab for their guidance and expertise. The work was supported by the Bass Connections Student Research Award for the year 2024-2025, Duke University.

\bibliographystyle{IEEEtran}

\bibliography{bibtex}

\end{document}